\documentstyle[12pt]{article}
\newcommand{\be}{\begin{equation}}
\newcommand{\ee}{\end{equation}}
\newcommand{\bea}{\begin{eqnarray}}
\newcommand{\eea}{\end{eqnarray}}
\newcommand{\ba}{\begin{array}}
\newcommand{\ea}{\end{array}}
\newcommand{\bmat}{\left(\ba}
\newcommand{\emat}{\ea\right)}

\newcommand{\norsl}{\normalsize\sl}
\newcommand{\norsc}{\normalsize\sc}
\begin{document}

\begin{titlepage}

\title
{Spin Structure Function $g_2(x,Q^2)$ \\ and \\ Twist-3 Operators
in QCD}

\author{
\norsc  Jiro KODAIRA\thanks{Supported in part by
          the Monbusho Grant-in-Aid for Scientific Research
          No. C-05640351.}\, and
           Yoshiaki YASUI\thanks{Supported in part by
          the Monbusho Grant-in-Aid for Scientific Research
          No. 050076}           \\
\norsl  Dept. of Physics, Hiroshima University\\
\norsl  Higashi-Hiroshima 724, JAPAN\\
\\
\norsc  Tsuneo UEMATSU\thanks{Supported in part by
           the Monbusho Grant-in-Aid for Scientific Research
          No. C-06640392 and \quad No. 06221239}\\
\norsl  Dept. of Fundamental Sciences\\
\norsl  FIHS, Kyoto University\\
\norsl  Kyoto 606-01, JAPAN}

\date{}
\maketitle

\begin{abstract}
{\normalsize
We investigate the spin structure function $g_2 (x, Q^2 )$
in the framework of the operator product expansion and the
renormalization group. The twist-3 operators appearing in
QCD are examined and their relations are studied.
It is noted that operators proportional to equation of motion
appear in the operator mixing through renormalization, which
can be studied from the relevant Green's functions.
We also note that
the coefficient functions can be properly fixed after the choice
of independent operators.
}
\end{abstract}

\begin{picture}(5,2)(-320,-555)
\put(2.3,-65){KUCP-71}
\put(2.3,-80){HUPD-9411}
\put(2.3,-95){August, 1994}
\end{picture}

\thispagestyle{empty}
\end{titlepage}
\setcounter{page}{1}
\baselineskip 24pt

Recently there has been much interest in the nucleon spin
structure functions which can be measured by the polarized
deep inelastic leptoproduction.
The nucleon spin structure is described by the two spin
structure functions $g_1(x,Q^2)$ and $g_2(x,Q^2)$.
The data from the EMC Collaboration~\cite{EMC} on $g_1$ have
prompted many authors to reanalyze the QCD effects on this process
and recent data from SMC at CERN~\cite{SMC} and E142 at
SLAC~\cite{E142} have attracted
much more attention in connection with the Bjorken sum rule.
On the other hand, the experiment on $g_2$ is expected to be
performed at CERN, SLAC and DESY in the near future.

For $g_1$, the leading contribution comes only from twist-2
operators in the operator product expansion (OPE), while
for $g_2$ the twist-3 operators also contribute
in the leading order of $1/Q^2$ in addition to the twist-2
operators~\cite{HM}.
This leads to new features which do not appear in the
analyses of other structure functions.

After the old papers~\cite{ARS,KOD1,KOD2}
which discussed the QCD effects on the polarized process,
the above problems have been addressed by many
authors~\cite{SVETAL,JAFFE}. It turns out that the procedure to
obtain the QCD corrections to the Wilson's coefficient functions of
each operator is not so straightforward for $g_2$.
The presence of the operators proportional to
the {\norsl equation of motion} brings about
some complication for the twist-3 operators,
through the course of renormalization, which is a general feature
characteristic to the higher-twist operators~\cite{POLI}.

In this paper we focus our attention on
the structure function $g_2(x,Q^2)$,
and study twist-3 operators in QCD in the framework of OPE
and renormalization group.
We examine the operator mixing under the renormalization
and point out that the coefficient functions
depend upon the choice of the basis of the independent operators.

Let us start with the basic formalism based on the operator
product expansion for analyzing the polarized
leptoproduction~\cite{KOD1}.
The hadronic tensor $W_{\mu\nu}$ is given by the absorptive part
of the forward virtual Compton amplitude. It's anti-symmetric part
$W_{\mu\nu}^{A}$ can be expressed by the two
spin-dependent structure functions $g_1$ and $g_2$ as :
\[
  W^A_{\mu\nu} = \varepsilon _{\mu\nu\lambda\sigma} q^{\lambda}
                \left\{ s^{\sigma} \frac{1}{p\cdot q} g_1(x, Q^2 )
                + ( p\cdot q s^{\sigma} - q\cdot s p^{\sigma} )
                 \frac{1}{(p\cdot q)^2} g_2 (x, Q^2 )\right\},
\]
where
$q$ is the virtual photon momentum and $p$ is the nucleon
momentum.
$x$ is the Bjorken variable $x=Q^2/2p\cdot q=Q^2/2M\nu$ ,
$p\cdot q=M\nu$ and $q^2=-Q^2$. $M$ is the nucleon mass and
$s^{\mu}={\bar u}(p,s)\gamma^{\mu}\gamma_5u(p,s)$ is the covariant
spin vector.

Applying OPE to the product of two
electromagnetic currents, we get for the anti-symmetric part,
\bea
  i\int d^4xe^{iq\cdot x} T(J_{\mu}(x)J_{\nu}(0))^A
      &=& -i\varepsilon_{\mu\nu\lambda\sigma} q^{\lambda}
         \sum_{n=1,3,\cdots} \left( \frac{2}{Q^2 }\right )^n
            q_{\mu _1} \cdots q_{\mu _{n-1}} \nonumber\\
    & &  \qquad \times \Bigl \{ E_q^n R_q^{\sigma\mu_{1}\cdots \mu_{n-1}}
           + \sum_j E_j^n R_j^{\sigma\mu_{1}\cdots \mu_{n-1}} \Bigr \} ,
\label{cpope}
\eea
where $R_i^n$'s are the composite operators and $E_i^n$'s
the corresponding coefficient functions.
In (\ref{cpope}), $R_q$ represents the twist-2 operators
and the other operators inside the summation over $j$
are the twist-3 operators. For simplicity, let us consider
the flavor non-singlet case.
(In the following expressions,
we suppress the flavor matrices $\lambda_i$ for the quark field
$\psi$ )
The explicit forms of the twist-2 operators are given by
\[
  R_q^{\sigma\mu_{1}\cdots \mu_{n-1}} =
         i^{n-1} \overline{\psi}\gamma_5 \gamma^{\{\sigma}D^{\mu_1}
              \cdots D^{\mu_{n-1}\}}\psi \ -{\rm (traces)} \ ,
\]
where $\{ \quad \}$ means symmetrization over the Lorentz indices
and $-$ (traces) stands for the subtraction of the trace terms to
make the operators traceless, which will be suppressed
in the following.
For the twist-3 operators, we have
\bea
  R_F^{\sigma\mu_{1}\cdots \mu_{n-1}} &=&
         \frac{i^{n-1}}{n} \Bigl[ (n-1) \overline{\psi}\gamma_5
       \gamma^{\sigma}D^{\{\mu_1} \cdots D^{\mu_{n-1}\}}\psi
\nonumber\\
   & & \qquad\qquad - \sum_{l=1}^{n-1} \overline{\psi} \gamma_5
       \gamma^{\mu_l }D^{\{\sigma} D^{\mu_1} \cdots D^{\mu_{l-1}}
            D^{\mu_{l+1}} \cdots D^{\mu_{n-1}\}}
                             \psi \Bigr] , \label{quark}\\
  R_m^{\sigma\mu_{1}\cdots \mu_{n-1}} &=&
          i^{n-2} m \overline{\psi}\gamma_5
       \gamma^{\sigma}D^{\{\mu_1} \cdots D^{\mu_{n-2}}
        \gamma ^{\mu_{n-1}\}} \psi , \label{mass} \\
  R_k^{\sigma\mu_{1}\cdots \mu_{n-1}} &=& \frac{1}{2n}
              \left( V_k - V_{n-1-k} + U_k + U_{n-1-k} \right) ,
                                                           \label{gluon}
\eea
where $m$ in (\ref{mass}) represents the quark mass (matrix).
The operators in (\ref{gluon}) contain
the gluon field strength $G_{\mu\nu}$ and the dual tensor
$\widetilde{G}_{\mu \nu}={1\over
2}\varepsilon_{\mu\nu\alpha\beta}
G^{\alpha\beta}$ explicitly and given by
\bea
    V_k &=& i^n g S \overline{\psi}\gamma_5
       D^{\mu_1} \cdots G^{\sigma \mu_k } \cdots D^{\mu_{n-2}}
        \gamma ^{\mu_{n-1}} \psi , \nonumber \\
    U_k &=& i^{n-3} g S \overline{\psi}
       D^{\mu_1} \cdots \widetilde{G}^{\sigma \mu_k } \cdots
             D^{\mu_{n-2}} \gamma ^{\mu_{n-1}} \psi , \nonumber
\eea
where $S$ means the symmetrization over $\mu_i$ and $g$ is the QCD
coupling constant.

It is by now well-known~\cite{SVETAL,JAFFE} that
the operators (\ref{quark} - \ref{gluon}) are
related through the operators which are proportional
to the {\norsl equation of motion} ( EOM operators ),
\bea
     R_{eq}^{\sigma\mu_{1}\cdots \mu_{n-1}} &=&
           i^{n-2} \frac{n-1}{2n} S [ \overline{\psi} \gamma_5
          \gamma^{\sigma} D^{\mu_1} \cdots D^{\mu_{n-2}}
        \gamma ^{\mu_{n-1}} (i\not{\!\!D} - m )\psi \nonumber\\
    & & \qquad\qquad\qquad\qquad + \overline{\psi} (i\not{\!\!D} - m )
              \gamma_5 \gamma^{\sigma} D^{\mu_1} \cdots D^{\mu_{n-2}}
        \gamma ^{\mu_{n-1}} \psi ] \nonumber \,.
\eea
It is not difficult to obtain the following relation :
\be
     R_F^{\sigma\mu_{1}\cdots \mu_{n-1}} =
        \frac{n-1}{n} R_m^{\sigma\mu_{1}\cdots \mu_{n-1}}
             + \sum_{k=1}^{n-2} (n-1-k)
                 R_k^{\sigma\mu_{1}\cdots \mu_{n-1}} +
             R_{eq}^{\sigma\mu_{1}\cdots \mu_{n-1}} .
\label{oprelation}
\ee
This equation means that these operators are not all independent
but
constrained by this relation, and therefore the renormalization
procedure
for the twist-3 operators becomes rather complicated.
Although we realize that physical matrix elements of the EOM
operators vanish~\cite{POLI}, we keep them to study the operator
mixing through renormalization
because their off-shell Green's
function do not vanish and are relevant for our analyses shown
below.

The anomalous dimensions which enter into
the renormalization
group equation for the coefficient functions,
are obtained from the renormalization
constant for the composite operators.
The anomalous dimensions of twist-3 operators in QCD at the
one-loop level have been studied in ref.\cite{SVETAL}.
Here we reanalyze the operator mixing problems by keeping
the EOM operators. We examine what happens to the
renormalization if there
are several operators which are related by constraints.
Let us first illustrate the points with the scalar field theory for
simplicity.
Assume that we have two composite operators $R_1$ and $R_2$
which are related through the equation of motion:
\be
R_1=R_2+E,
\label{constr}
\ee
where $E$ is a EOM operator.
The general form of $E$ will be
$E = A(\phi )\,\,\delta S /\delta \phi $ where $S$ is the
action and $A(\phi)$ is a some function of $\phi$.
Since we have three composite operators with a
constraint~(\ref{constr}),
we can choose any two operators among three as an independent basis.

First we choose $R_1$ and $R_2$. Then renormalization takes the
following form:
\be
\bmat{c}
R_1 \\ R_2
\emat_{\hspace{-0.1cm}R}
=
\bmat{cc}
Z_{11} & Z_{12} \\
Z_{21} & Z_{22}
\emat
\bmat{c}
R_1 \\ R_2
\emat_{\hspace{-0.1cm}B}.
\label{rerr}
\ee
Here the suffix $R$ ($B$) denotes the renormalized (bare) operator.
Next,
we change the operator basis to $R_i$ ($i=1$ or 2) and $E$. Then the
renormalization matrix becomes triangular~\cite{COLL},
\be
\bmat{c}
R_i \\ E
\emat_{\hspace{-0.1cm}R}
=
\bmat{cc}
Z_{ii}^E & Z_{iE} \\
0 & Z_{EE}
\emat
\bmat{c}
R_i \\ R_E
\emat_{\hspace{-0.1cm}B},
\label{rere}
\ee
since the counterterm to the operator $E$ should vanish by the
equation of motion.  From (\ref{rerr}) and (\ref{rere}), we can derive
the following relations between the renormalization constants
$Z$'s:
\bea
Z_{11}^E = Z_{11}+Z_{12} &=& Z_{21}+Z_{22}=Z_{22}^E
,\label{zrel}\\
Z_{EE}=Z_{22}-Z_{12}=Z_{11}-Z_{21}\,,& &
Z_{1E}=-Z_{12}\,,\qquad Z_{2E}=Z_{21} .\nonumber
\eea
Eq.(\ref{zrel}) is a consistency condition for the renormalization
of the operators satisfying (\ref{constr}).

Here we note
that the arbitrariness in the choice of the operator basis does not
enter into the physical quantities.
What we really need is the physical matrix element of operators.
This fact can be explicitly confirmed by noting that the physical
matrix element of the operator $E$
vanishes~\cite{POLI,COLL}:\,\, $\langle p |E|p'\rangle =0.$
Taking the physical matrix element of (\ref{rere}), we conclude
that the renormalization constant $Z_{ii}^E$ is only relevant.
We reach the same conclusion also from (\ref{rerr}) by taking
account of the fact:
$ \langle p |R_1|p'\rangle =\langle p |R_2|p'\rangle $
and (\ref{zrel}).
The explicit one loop calculation shows that for the simplest
twist-3 operator ${\bar \psi}D_{\mu}D_{\nu}\psi$, as discussed
by Jaffe \cite{JAFFE}, the above argument actually holds \cite{KU}.

Now we go back to the case of polarized deep inelastic scattering
and give explicit results at the one-loop level for
$n=3$ as an example. In this case we have
four operators with the constraint (\ref{oprelation}).
\be
  R_F = \textstyle{\frac{2}{3}} R_m + R_1 + R_{eq},
\label{n3}
\ee
where Lorentz indices of operators are suppressed.
Now we choose $R_F, R_m, {\rm and}R_1$ as independent operators
and eliminate EOM operator $R_{eq}$. In order to renormalize
above operators, it turns out that a gauge non-invariant EOM
operator $R_{eq1}$ comes into play,
\[
     R_{eq1}^{\sigma\mu_{1} \mu_{2}} =
           i \textstyle{1\over 3} S [ \overline{\psi} \gamma_5
          \gamma^{\sigma} \partial ^{\mu_1}
        \gamma ^{\mu_{2}} (i\not{\!\!D} - m )\psi +
      \overline{\psi} (i\not{\!\!D} - m )
              \gamma_5 \gamma^{\sigma} \partial ^{\mu_1}
        \gamma ^{\mu_{2}} \psi ] .\]
Its presence is allowed because it vanishes by  the equation
of motion~\cite{COLL}.
We found the following renormalization
constant for the composite operators,
\be
\bmat{c}
R_F \\ R_1 \\ R_m \\ R_{eq1}
\emat_{\hspace{-0.1cm}R}
=
\bmat{cccc}
Z_{11} & Z_{12} & Z_{13} & Z_{14} \\
Z_{21} & Z_{22} & Z_{23} & Z_{24} \\
0 & 0 & Z_{33} & 0 \\
0 & 0 & 0 & Z_{44}
\emat
\bmat{c}
R_F \\ R_1 \\ R_m \\ R_{eq1}
\emat_{\hspace{-0.1cm}B}
\label{n3z1}
\ee
where
$Z_{ij}$ are given in the
dimensional regularization
$D=4-2\varepsilon$:
\[
Z_{ij}\equiv \delta_{ij} +
{1\over\varepsilon}{{g^2}\over{16\pi^2}}z_{ij}
\]
with
\be
\ba{ll}
z_{11}={7\over 6}C_2(R)+{3\over 8}C_2(G), &
z_{12}=-{3\over 2}C_2(R)+{21\over 8}C_2(G), \\
z_{13}=3C_2(R)-{1\over 4}C_2(G), & z_{14}=-{3\over 8}C_2(G), \\
z_{21}={1\over 6}C_2(R)-{1\over 8}C_2(G), &
z_{22}=-{1\over 2}C_2(R)+{25\over 8}C_2(G), \\
z_{23}=-{1\over 3}C_2(R)+{1\over 12}C_2(G), &
z_{24}={1\over 8}C_2(G), \\
z_{33}=6C_2(R), & z_{44}=0 .
\label{zn3}
\ea
\ee
The quadratic Casimir operators are $C_2(R)=4/3$ and $C_2(G)=3$
for the case of QCD.
In the above calculations, it is necessary to compute the
{\norsl off-shell Green's functions} of the
composite operators. Otherwise some informations on the
renormalization constants associated with the EOM
operator will be lost.

As explained before, one can choose other operators as an
independent
basis using (\ref{n3}). The consistency conditions corresponding
to (\ref{zrel}) read in this case,
\[
\ba{ll}
z_{11}+z_{12} = z_{21}+z_{22}, &
    {2\over 3}z_{11}+z_{13} = {2\over 3}z_{21}+z_{23}+{2\over 3}z_{33}, \\
  z_{13}-{2\over 3}z_{12} = z_{23}-{2\over 3}z_{22}+{2\over 3}z_{33}.&
\ea
\]
These equalities are indeed satisfied with (\ref{zn3}).

The authors in ref.\cite{SVETAL} have discarded the fermion
bilinear operator $R_F^{\sigma\mu_{1}\cdots \mu_{n-1}}$ in their
analyses.
For the present case of $n=3$, the renormalization constant
matrix becomes
\be
\bmat{c}
R_1 \\ R_m \\ R_{eq} \\ R_{eq1}
\emat_{\hspace{-0.1cm}R}
=
\bmat{cccc}
Z_{21}+Z_{22} & {2\over 3}Z_{21}+Z_{23} & Z_{21} & Z_{24} \\
0 & Z_{33} & 0 & 0 \\
0 & 0 & Z_{11}-Z_{21} & Z_{14}-Z_{24} \\
0 & 0 & 0 & Z_{44}
\emat
\bmat{c}
R_1 \\ R_m \\ R_{eq} \\ R_{eq1}
\emat_{\hspace{-0.1cm}B} ,
\ee
where
$Z_{ij}$ are defined in (\ref{n3z1}). In this basis, our
results agree with those in ref.\cite{SVETAL}.
Although this choice of basis might be economical especially for
general $n$, it is never
compulsory. One can eliminate any operator as well.
Any choice of basis leads to a unique prediction
for the moment with the properly determined
coefficient functions discussed below.

To make a prediction for the moments, we must determine
the coefficient functions corresponding to the \lq\lq hard
part\rq\rq  of the process.
The coefficient functions at the tree level can be obtained by
considering
the short distance expansion of the current product in the presence
of external gauge field $A_{\mu}^a$. Up to twist-3 operators, the
OPE reads~\cite{SVETAL,JAFFE}
\bea
 & & i\int d^4xe^{iq\cdot x} T(J_{\mu}(x)J_{\nu}(0))^A \nonumber\\
 & & \qquad = - i\varepsilon_{\mu\nu\lambda\sigma} q^{\lambda}
         \sum_{n=1,3,\cdots} \left( \frac{2}{Q^2 }\right )^n
            q_{\mu _1} \cdots q_{\mu _{n-1}}
       \bigl \{ R_q^{\sigma\mu_1\cdots\mu_{n-1}} + R_F^{\sigma\mu_1
                          \cdots\mu_{n-1}} \bigr \} .\label{extope}
\eea
Here if we use (\ref{oprelation}),
$R_F^{\sigma\mu_1\cdots\mu_{n-1}}$ can
be eliminated in terms of other operators.
{}From this observation, it is inferred
that the (tree-level) coefficient functions depend upon the
choice of the independent operators. In the
basis of independent operators which includes $R_F^n$,
we conclude that at the tree-level,
\be
  E_q^n ({\rm tree})= E_F^n ({\rm tree}) = 1\,,\qquad
    E_m^n ({\rm tree}) = E_j^n ({\rm tree}) = 0 \,.
\label{coef1}
\ee
On the other hand, if we eliminate $R_F^n$ we have,
\be
  E_q^n ({\rm tree})=1\,,\quad E_m^n ({\rm tree})= {{n-1}\over n}\,,
   \quad E_j^n ({\rm tree}) = n-1-j.
\label{coef2}
\ee

Now we shall write down the moment sum rule for $g_2$.
Define the matrix elements of composite operators between nucleon
states with momentum $p$ and spin $s$ by
\bea
  \langle p,s | R_q^{\sigma\mu_{1}\cdots \mu_{n-1}} |p,s \rangle
       &=& - a_n s^{\{\sigma}p^{\mu_1} \cdots p^{\mu_{n-1}\}}
                           \label{element1}\\
  \langle p,s | R_F^{\sigma\mu_{1}\cdots \mu_{n-1}} |p,s \rangle
       &=& -  \frac{n-1}{n} d_n ( s^{\sigma}p^{\mu_1} - s^{\mu_1}p^{\sigma})
                    p^{\mu_2} \cdots p^{\mu_{n-1}} \\
  \langle p,s | R_m^{\sigma\mu_{1}\cdots \mu_{n-1}} |p,s \rangle
       &=& - e_n ( s^{\sigma}p^{\mu_1} - s^{\mu_1}p^{\sigma})
                    p^{\mu_2} \cdots p^{\mu_{n-1}} \\
  \langle p,s | R_k^{\sigma\mu_{1}\cdots \mu_{n-1}} |p,s \rangle
       &=& - f_n^k ( s^{\sigma}p^{\mu_1} - s^{\mu_1}p^{\sigma})
                    p^{\mu_2} \cdots p^{\mu_{n-1}} \\
  \langle p,s | R_{eq}^{\sigma\mu_{1}\cdots \mu_{n-1}} |p,s \rangle
      &=& 0 \label{elementeq}.
\eea
Our normalization in the above definition is such that for free
quark target $a_n = d_n = e_n = 1$. On the other hand, $f_n^k
= {\cal O} (g^2 )$. Using (\ref{element1} - \ref{elementeq}),
we can write down the moment sum rule for $g_2$.
\bea
    M_n \equiv \int_0^1 dx x^{n-1} g_2 (x,Q^2) &=&
          - {{n-1}\over {2n}} \Bigl [ a_n E_q^n(Q^2)
              - d_n E_F^n(Q^2) \Bigr ] \nonumber\\
    & & \qquad\qquad + {1\over 2} \Bigl [ e_n E_m^n(Q^2)
         + \sum_{j=1}^{n-2}f_n^jE_j^n(Q^2) \Bigr ].
\label{g2sumrule}
\eea
with the following constraint from (\ref{oprelation}),
\[
    {{n-1}\over n}d_n={{n-1}\over n}e_n+\sum_{j=1}^{n-2}(n-1-j)f_n^j.
\]
It should be stressed that the explicit form of the $Q^2$ evolution
of each term in (\ref{g2sumrule}) depends on the choice of operator
basis. The anomalous dimensions as well as the coefficient
functions
$E_j^n$ take different forms depending on the basis. However
the moment itself remain the same as one expects. For the case
of $n=3$, we have explicitly checked that any choice of the basis
leads to the same expression for the moment.

Here we note that the coefficient functions for the EOM
operators do not mix with those for the other operators.
This is due to the triangular structure of the anomalous
dimension matrix in their renormalization-group equation.
This fact is essential to the equivalence of the moments for
any basis.

Let us now consider the moments of $g_2$ in the basis of $R_j$ and $R_m$.
In this case,
\be
    M_n = - {{n-1}\over {2n}} a_n E_q^n(Q^2)
    + {1\over 2} \Bigl [ e_n E_m^n(Q^2)
         + \sum_{j=1}^{n-2}f_n^jE_j^n(Q^2) \Bigr ].
\label{moment}
\ee
with the coefficient functions given in (\ref{coef2}).
We shall show below that (\ref{moment}) indeed holds in the leading
order of $\ln{Q^2}$.

The operator mixing of gluon-field dependent operators $R_j$ and
the mass dependent operator $R_m$ through the renormalization reads
neglecting the EOM operators,
\[
\bmat{c}
R_j \\ R_m
\emat_{\hspace{-0.1cm}R}
=
\bmat{cc}
Z_{ji} & Z_{jm} \\
0 & Z_{mm}
\emat
\bmat{c}
R_i \\ R_m
\emat_{\hspace{-0.1cm}B}.
\]
By evaluating the Green's functions of these operators with
incoming and outgoing off-shell quark states, we obtain the off-diagonal
element of the renormalization constant matrix $Z_{jm}$ from which
the anomalous dimension reads,
\be
\gamma_{mj}\equiv -{{g^2}\over {16\pi^2}}{{8C_2(R)}\over n}
{1\over{j(j+1)(j+2)}}\equiv {{g^2}\over {16\pi^2}}\gamma_{mj}^0.
\label{gammj}
\ee
This result is in disagreement with the one given in the fifth
reference in \cite{SVETAL}.

Now note that the Compton scattering amplitude off the on-shell
massive quark target has been calculated~\cite{KOD1},
in the leading order of $\ln{Q^2}$, to be
\be
    M_n = {1\over 2}{g^2\over{16\pi^2}}C_2(R)(-2+{4\over{n+1}})\ln{Q^2}
                   +\cdots .
\label{comp}
\ee
Whereas, using the perturbative solution for the
renormalization-group equation,
the right-hand side of (\ref{moment}) becomes
for the quark matrix elements,
\be
{\rm RHS \ of \ }(\ref{moment}) = - {1\over 2}{g^2\over{16\pi^2}}
      \left ( {{n-1}\over n}\Bigl( -{1\over 2}\Bigr) \gamma_q^0
            E_q^n ({\rm tree})+{1\over 2}\gamma_{mj}^0
                 E_j({\rm tree}) \right ) \ln{Q^2} + \cdots .
\label{g2ope}
\ee
Putting $\gamma_{mj}^0$ of (\ref{gammj}) and the following
expression for the anomalous dimensions,
\[
\gamma_q^0 = 2C_2(R) \Bigl [ 1-{2\over{n(n+1)}} +
   4\sum_{j=2}^n{1\over j} \Bigr ],\qquad
          \gamma_{mm}^0=8C_2(R) \sum_{j=1}^{n-1}{1\over j}.
\]
into the above equation
with the tree level coefficient functions (\ref{coef2}), we find
(\ref{g2ope}) coincides with (\ref{comp}).
Thus we confirm that our results (\ref{gammj},\ref{coef2})
are consistent with the moment sum rule.

Finally let us comment on the Burkhardt-Cottingham sum rule
which corresponds to the first moment of $g_2$.
We see from (\ref{quark}-\ref{gluon}) that
twist-3 operators can {\sl not} be defined for $n=1$.
Therefore the OPE analysis suggests that
the Burkhardt-Cottingham sum rule:
\[
\int_0^1dx \ g_2(x,Q^2)=0
\]
does not receive any corrections in QCD~\cite{ALNR,KMSU}.

To summarize, we have  examined the twist-3 operators
in QCD which contribute to $g_2(x,Q^2)$.
In the renormalization, there appear
operators proportional to the equation of motion, which can
be studied by computing off-shell Green's functions.
Because of the relationship among the twist-3 operators,
we have to choose a basis of independent operators, by which
the coefficient functions are properly fixed.
We have also noted that the Burkhardt-Cottingham sum rule
holds in the higher-order of QCD ~\cite{KMSU}.

We expect that future measurements on $g_2$ will clarify the
effect of twist-3 operators which share some common feature
with more general higher-twist effects in QCD.

\vspace{1cm}
\noindent
We would like to thank Shoji Hashimoto and Ken Sasaki for
discussions.


\end{document}